\documentclass[
]{ceurart}

\usepackage{listings}
\usepackage{multirow}
\usepackage[normalem]{ulem}
\useunder{\uline}{\ul}{}

\begin{document}

\copyrightyear{2024}
\copyrightclause{Copyright for this paper by its authors.
  Use permitted under Creative Commons License Attribution 4.0
  International (CC BY 4.0).}

\conference{CLEF 2024: Conference and Labs of the Evaluation Forum, September 09–12, 2024, Grenoble, France}

\title{A Time-Aware Approach to Early Detection of Anorexia: UNSL at eRisk 2024}

\author[1,2]{Horacio Thompson}[%
email=hjthompson@unsl.edu.ar,
]

\address[1]{Universidad Nacional de San Luis (UNSL), Ejército de Los Andes 950, San Luis, C.P. 5700, Argentina}

\address[2]{Consejo Nacional de Investigaciones Científicas y Técnicas (CONICET), San Luis, Argentina}

\author[1]{Marcelo Errecalde}[%
email=merreca@unsl.edu.ar,
]

\begin{abstract}
The eRisk laboratory aims to address issues related to early risk detection on the Web. In this year's edition, three tasks were proposed, where Task 2 was about early detection of signs of anorexia. Early risk detection is a problem where precision and speed are two crucial objectives. Our research group solved Task 2 by defining a \emph{CPI+DMC approach}, addressing both objectives independently, and a \emph{time-aware approach}, where precision and speed are considered a combined single-objective. We implemented the last approach by explicitly integrating time during the learning process, considering the ERDE$_{\theta}$ metric as the training objective. It also allowed us to incorporate temporal metrics to validate and select the optimal models. We achieved outstanding results for the ERDE$_{50}$ metric and ranking-based metrics, demonstrating consistency in solving ERD problems.
\end{abstract}


\maketitle

\section{Introduction}

According to the World Health Organization, approximately one in every eight people worldwide suffers from a mental disorder, with anxiety, depression, bipolar disorder, and eating disorders being the most prevalent \cite{charlson2019new}. Multiple studies have underscored the correlation between social media usage and mental health disorders \cite{aliverdi2022social, maulik2010effect, martinez2022redes, nawaz2024social}. Early Risk Detection (ERD) on the Web consists of correctly identifying users who show signs of mental health conditions as soon as possible. The Early Risk Prediction on the Internet (eRisk) laboratory has addressed challenges related to ERD problems in its different editions. This year, three tasks were proposed \cite{parapar2024, parapar2024ext}, Task 2 being about early detection of signs of anorexia.

ERD incorporates a significant complexity to standard classification problems since the users are analyzed post-by-post rather than processing the complete history. In recent editions of eRisk, our research group proposed solutions based on a \emph{CPI+DMC approach}, considering ERD as a multi-objective problem. The goal is to find an optimal balance between correctly identifying at-risk users and minimizing the time needed to make reliable decisions, addressing precision and speed independently. It involves defining two components: one dedicated to solving a classification problem with partial information (CPI) and the other to deciding the moment of classification (DMC). Using this approach, we achieved interesting results in the 2021 \cite{loyola2021unsl}, 2022 \cite{loyola2022unsl}, and 2023 \cite{thompson2023strategies} eRisk editions. In the last two editions, we applied the BERT model \cite{Devlin2018} with an extended vocabulary for the CPI component and a decision policy based on the model’s prediction history during user evaluation (\emph{historic rule}) for the DMC component. Similarly, in \cite{thompson2023early}, we applied these methods in MentalRiskES 2023, the first ERD challenge for the Spanish language.
In this case, we used the BETO model \cite{CaneteCFP2020}, a variant of BERT for Spanish, and adjusted the \emph{historic rule} according to the tasks to be solved, obtaining excellent results.
  
Although precision is crucial for ERD problems, as time progresses and decisions are delayed, speed becomes as important as precision. In this context, it could be proper to consider ERD as a combined single-objective problem, simultaneously considering precision and speed within the learning process. Proposals such as \cite{loyola2021unsl, loyola2022unsl} applied the EARLIEST architecture for time series \cite{earliest} in ERD, aiming to balance both objectives through reinforcement learning. Since the advent of transformers \cite{Vaswani2017}, numerous studies have focused on enhancing user classification through the standard fine-tuning process, but often without addressing speed as a crucial aspect in decision-making.
Models are trained and validated differently than how they are evaluated in ERD challenges, making it difficult to select the optimal model. 
Therefore, we propose a \emph{time-aware approach}, which involves applying the fine-tuning process considering the user information and time progress. 
By incorporating time into the learning process, it is possible to define a scenario similar to that used in the evaluation stage in ERD.
This allows us to select the optimal model using temporal metrics such as ERDE$_{\theta}$ while avoiding the necessity of defining complex decision policies for an early detection environment. 

Considering both approaches, we addressed Task 2 of the eRisk 2024 edition using three models, two of which implemented the \emph{CPI+DMC approach}, while the third adopted the \emph{time-aware approach}. We achieved the second-best score for the ERDE$_{50}$ metric and the first-best place in several rankings-based metrics, as well as an acceptable performance compared to the mean among all teams for the F$_{1}$ and F$_{latency}$ metrics. Additionally, we were the second-fastest team to solve the proposed task. 
Therefore, our proposals demonstrated consistency when evaluated for the early detection of signs of anorexia.

\section{Time-aware approach}

We propose a method that tunes transformer-based models by incorporating time during the learning process. The process involves modifying the inputs by adding time, applying the fine-tuning process by defining ERDE$_{\theta}$ as the training objective, and then evaluating the selected best model in an ERD environment.

\medskip \noindent
\textbf{Data preprocessing.} Time is explicitly included in the input samples, considering the number of posts that have been read until the moment (\emph{delay}). For an arbitrary sample $\langle$input, label$\rangle$, the new input is defined as input $_{delay}$ = [CLS] input [TIME]  $delay$  [SEP]. The [TIME] token is added to the model architecture, separating the post content of the moment it was read. For example, \emph{I don't feel like eating} evaluated at  $delay=10$, the model would receive [CLS] \emph{I don't feel like eating} [TIME] $10$ [SEP].

\medskip \noindent
\textbf{Time-aware fine-tuning.} The process is carried out in epochs with the typical training and validation stages. 
We propose to incorporate an evaluation scheme similar to the testing environment for ERD problems in both stages, where time is measured based on \emph{delays}.
At each \emph{delay}, post windows of length M are evaluated by concatenating the current post with the previous M-1 posts. The \emph{delays} are configured based on the window size. For instance, with M=10, \emph{delay}=10 assesses posts 0 to 9, \emph{delay}=20 assesses posts 10 to 19, and so forth until all users have been analyzed.
The \emph{loss} function is computed at the end of each \emph{delay}, evaluating the post window for users still undergoing analysis, and then the gradient is propagated throughout the transformer. We used the ERDE$_{\theta}$ metric \cite{losada2016} to design a linear and differentiable \emph{loss} function. This involved incorporating classification performance (CrossEntropy) and heavily penalizing delayed true positives based on a threshold $\theta$. Thus, as the \emph{loss} is minimized, ERDE$_{\theta}$ is also reduced, establishing it as the training objective. The validation stage follows a similar \emph{delay} scheme, concluding each epoch with the calculation of \emph{loss}, \emph{accuracy}, and ERDE$_{\theta}$ metrics. These metrics can be weighted to select the optimal model for an ERD problem. 

\medskip \noindent
\textbf{Model testing in ERD.} The best model obtained can be evaluated using a \emph{mock-server} tool\footnote{Available at: \url{https://github.com/jmloyola/erisk\_mock\_server}.}, which simulates ERD environments through rounds of posts and answers submissions and calculates the results using different metrics. A client application was defined to interact with the server: when it receives a round of posts, the system preprocesses them by adding time, invokes the predictive model, and returns a response. We used a sliding post window, configured as in the learning stage, and a simple decision policy (\emph{simple rule}): if the probability exceeds a \emph{threshold} (the prediction limit probability for a positive user), a user at-risk alarm is issued; otherwise, the analysis should continue.

\section{Task resolution}
Task 2 was conducted following the guidelines from previous editions. It consisted of two stages: a \emph{training stage}, where participants experimented with the data provided by the organizers, followed by a \emph{testing stage}. During the latter, an early environment was established, wherein each participant deployed a client application to interact with a server, retrieving user posts one by one and submitting responses using the proposed predictive models. In this section, we provide details of the datasets involved in the task, the models proposed by our team, and the results achieved.

\subsection{Datasets}
Table \ref{tab:ds} shows the details of the corpora available to solve the task. The organizers used the eRisk2024 corpus to evaluate the participating models, while the other corpora were released for participants to train their models. In our case, we trained the models using the eRisk2019 and eRisk2018\_train corpora, reserving the eRisk2018\_test to evaluate them in an early environment.
All datasets have an imbalance in the classes, with a similar proportion of positive users among them. 
In particular, the eRisk2024 corpus contains 784 users, with 12\% being positive, and compared to eRisk2018\_test, it has 40\% more samples, a greater number of posts per user, and more words per post.

\begin{table}[ht!]
\caption{Details of the corpora used to solve Task 2. The number of users (total, positives, and negatives) and the number of posts in each corpus are reported. The mean, minimum, and maximum number of posts per user and words per post in each corpus are detailed.}
\label{tab:ds}
\begin{tabular}{|l|ccc|ccc|ccc|}
\hline
\multicolumn{1}{|c|}{\multirow{2}{*}{\textbf{Corpus}}} & \multicolumn{3}{c|}{\textbf{\#users}}                                                  & \multicolumn{3}{c|}{\textbf{\#posts per user}}                                        & \multicolumn{3}{c|}{\textbf{\#words per post}}                                        \\ 
\multicolumn{1}{|c|}{}                                 & \multicolumn{1}{c|}{\textbf{Total}} & \multicolumn{1}{c|}{\textbf{Pos}} & \textbf{Neg} & \multicolumn{1}{c|}{\textbf{Mean}} & \multicolumn{1}{c|}{\textbf{Min}} & \textbf{Max} & \multicolumn{1}{c|}{\textbf{Mean}} & \multicolumn{1}{c|}{\textbf{Min}} & \textbf{Max} \\ \hline
eRisk2024                                              & \multicolumn{1}{c|}{784}            & \multicolumn{1}{c|}{92}           & 692          & \multicolumn{1}{c|}{626}           & \multicolumn{1}{c|}{10}           & 2,001        & \multicolumn{1}{c|}{35}            & \multicolumn{1}{c|}{1}            & 19,668       \\
eRisk2019                                              & \multicolumn{1}{c|}{815}            & \multicolumn{1}{c|}{73}           & 742          & \multicolumn{1}{c|}{700}           & \multicolumn{1}{c|}{10}           & 2,000        & \multicolumn{1}{c|}{32}            & \multicolumn{1}{c|}{1}            & 8,953        \\
eRisk2018\_train                                       & \multicolumn{1}{c|}{152}            & \multicolumn{1}{c|}{20}           & 132          & \multicolumn{1}{c|}{558}           & \multicolumn{1}{c|}{9}            & 1,999        & \multicolumn{1}{c|}{34}            & \multicolumn{1}{c|}{1}            & 7,404        \\
eRisk2018\_test                                        & \multicolumn{1}{c|}{320}            & \multicolumn{1}{c|}{41}           & 279          & \multicolumn{1}{c|}{527}           & \multicolumn{1}{c|}{9}            & 1,999        & \multicolumn{1}{c|}{33}            & \multicolumn{1}{c|}{1}            & 8,318        \\ \hline
\end{tabular}
\end{table}

\subsection{Models}
We presented three models to address Task 2, of which two applied the \emph{CPI+DMC approach}, and a third applied the \emph{time-aware approach}. Training and validation were performed by splitting the data in an 80/20 ratio. 
Several preprocessing steps were applied to the data before training, such as converting texts to lowercase, transforming Unicode and HTML codes into their corresponding symbols, replacing websites with the `weblink' token, and eliminating repeated words, among other steps. Subsequently, the best models were evaluated in an early environment using the \emph{mock-server} tool. Below are the details of each proposal.

\medskip \noindent
\textbf{UNSL\#0 - CPI+DMC approach} \\
\emph{CPI component.} Classic fine-tuning using the BERT model. Hyperparameters: Architecture=`BERT-based-uncased', optimizer=`AdamW', LR=3E-5, scheduler=`LinearSchedulerWarmup', batch\_size=8, and n\_epochs=5. These values were chosen based on the models' performance according to the F$_1$ metric on the positive class. \\
\emph{DMC component.} Decision policy based on the \emph{historic rule}: if the current prediction and last \emph{M} predictions exceed a \emph{threshold}, the client application must issue a risky user alarm; otherwise, the analysis should continue. In addition, the rule has the \emph{min\_delay} parameter, which defines the moment when it will begin to apply. Hyperparameters: \emph{threshold}=0.7, \emph{min\_delay}=10, \emph{M}=10. These values were obtained considering temporal metrics (ERDE$_{50}$ and F$_{latency}$) when evaluating in an early environment.

\medskip \noindent
\textbf{UNSL\#1 - CPI+DMC approach} \\
\emph{CPI component.} Classic fine-tuning using the BERT model with extended vocabulary. We applied the SS3 model \cite{burdisso2019text} to expand the BERT vocabulary, considering the most relevant words for the anorexia domain. We evaluated different models by varying the number of words, ultimately selecting the top 50 according to the confidence value of SS3 on the positive class. Thus, previously unknown words to BERT, such as \emph{calories}, \emph{binge}, \emph{tulpa}, \emph{purging}, \emph{vegan}, \emph{underweight}, and \emph{overweight}, were included. Additionally, we used the same hyperparameters as UNSL\#0, configuring LR=5E-5. \\
\emph{DMC component.} Decision policy based on the \emph{historic rule}, with the same hyperparameters as UNSL\#0.

\medskip \noindent
\textbf{UNSL\#2 - Time-aware approach} \\
We applied the \emph{time-aware} fine-tuning process by modifying the input texts and used ERDE$_{50}$ as the training objective since participants are usually evaluated using this metric. To configure \emph{delays} during the training and validation stages, we used a post window with a size of 10 and truncated the maximum length of user history to 200 posts to prevent bias in the model's learning. The best model was selected by weighting the accuracy and ERDE$_{50}$ metrics, obtaining the following hyperparameters: Architecture=`BERT-based-uncased', optimizer=`AdamW', LR=3E-5, scheduler=`LinearSchedulerWarmup', batch\_size=8, and n\_epochs=10. Furthermore, because the model learns the decision policy during training, we used a \emph{simple rule} configured with \emph{threshold}=0.7 and \emph{min\_delay}=10.

\subsection{Results}

In this section, we present the results obtained during the current edition of eRisk. A total of 44 proposals from 10 teams were submitted to solve Task 2 (Table \ref{tab:teams}).
Our team completed the task in 7 hours by processing all user posts through the three models previously described. This performance positioned us as the second-fastest team to complete the task.

\begin{table}[ht!]
\caption{Total time spent by each team for Task 2. The team name, number of models, and number of user posts processed are shown.}
\label{tab:teams}
\begin{tabular}{lrrr}
\hline
\textbf{Team} & \textbf{\#models} & \textbf{\#posts processed} & \textbf{Total time} \\ \hline
UMUTeam & 5 & 2001 & 6h:34m\\
UNSL & 3 & 2001 & 7h\\
BioNLP-IISERB & 5 & 10 & 9h:39m \\
NLP-UNED & 5 & 2001 & 9h:40m \\
ELiRF-UPV & 4 & 2001 & 12h:27m \\
Riewe-Perla & 5 & 2001 & 2 days + 11h:25m \\
GVIS & 5 & 352 & 3 days + 12h:36m \\
SINAI & 5 & 2001 & 3 days + 23h:49m \\
APB-UC3M & 2 & 2001 & 6 days 21h:34m \\
COS-470-Team-2 & 5 & 1 & - \\ \hline
\end{tabular}
\end{table}

Table \ref{tab:decision} displays the outcomes achieved by our team according to decision-based metrics. The three models attained the second-best ERDE$_{50}$, being surpassed by Riewe-Perla\#0, while NLP-UNED\#1 achieved top results in both F$_{1}$ and F$_{latency}$. 
Our models yielded satisfactory results, surpassing the mean of all teams for the F$_{1}$, ERDE$_{50}$, and F$_{latency}$ metrics. Moreover, UNSL\#2 (\emph{time-aware} model) exhibited the same performance as UNSL\#0 and UNSL\#1 (\emph{CPI+DMC} models) in the ERDE$_{50}$ metric. This fact underscores UNSL\#2's ability to optimize its ERDE$_{50}$ during training and apply a simple decision policy to solve the task, in contrast to the other models obtained through conventional fine-tuning, followed by a more complex policy.

\begin{table}[ht!]
\caption{Decision-based evaluation results for Task 2. The best teams taking into account the F$_{1}$, ERDE$_{5}$, ERDE$_{50}$, and F$_{latency}$ are shown, as well as the \emph{mean} and \emph{median} values of the results report for CLEF eRisk 2024. Values in bold and underlined depict 1st and 2nd performance achieved in the challenge, respectively.}
\label{tab:decision}
\begin{tabular}{|lcccccccc|}
\hline
\textbf{Model}  & \textbf{P}    & \textbf{R} & \textbf{F$_{1}$} & \textbf{ERDE$_{5}\downarrow$} & \textbf{ERDE$_{50}\downarrow$} & \textbf{latencyTP$\downarrow$} & \textbf{speed} & \textbf{F$_{latency}$} \\ \hline
UNSL\#0         & 0.35          & 0.99       & 0.52             & 0.14                          & {\ul 0.03}                     & 12                 & 0.96           & 0.49                   \\
UNSL\#1         & 0.42          & 0.96       & 0.59             & 0.14                          & {\ul 0.03}                     & 12                 & 0.96           & 0.56                   \\
UNSL\#2         & 0.42          & 0.97       & 0.59             & 0.14                          & {\ul 0.03}                     & 12                 & 0.96           & 0.56                   \\ \hline
Riewe-Perla\#0  & 0.45          & 0.97       & 0.62             & \textbf{0.07}                 & \textbf{0.02}                  & 6                  & 0.98           & 0.6                    \\
NLP-UNED\#1     & 0.67          & 0.97       & \textbf{0.79}    & 0.09                          & 0.04                           & 14                 & 0.95           & \textbf{0.75}          \\
\hline
\textit{Mean}   & 0.34          & 0.78       & 0.42             & 0.12                          & 0.07                           & 8.50               & 0.84           & 0.41                   \\
\textit{Median} & 0.41          & 0.97       & 0.49             & 0.11                          & 0.06                           & 6.00               & 0.96           & 0.47                   \\ \hline
\end{tabular}
\end{table}

Considering the ranking-based metrics (Table \ref{tab:ranking}), our team achieved the best results in multiple categories across different post counts (1, 100, 500, and 1000). 
The results were comparable to NLP-UNED\#1 and outperformed Riewe-Perla\#0 in all metrics. The only teams that showed acceptable results under these metrics were UNSL, NLP-UNED, and Riewe-Perla. In contrast, the overall performance of the other teams was considerably lower, as evidenced by the mean and median values. Additionally, UNSL\#1 and UNSL\#0 achieved better performance than UNSL\#2.

\begin{table}[ht!]
\footnotesize 
\caption{Ranking-based evaluation results for Task 2. Results are reported according to the three classification metrics obtained after processing 1, 100, 500, and 1000 posts, respectively. Values in bold represent the best performance achieved in the challenge.}
\label{tab:ranking}
\resizebox{\textwidth}{!}{%
\begin{tabular}{|l|ccc|ccc|ccc|ccc|}
\hline
\multicolumn{1}{|c|}{\textbf{}} & \multicolumn{3}{c|}{\textbf{1 post}} & \multicolumn{3}{c|}{\textbf{100 posts}} & \multicolumn{3}{c|}{\textbf{500 posts}} & \multicolumn{3}{c|}{\textbf{1000 posts}} \\
\textbf{Model} & \multicolumn{1}{c|}{\textbf{\rotatebox{90}{P@10} }} & \multicolumn{1}{c|}{\textbf{\rotatebox{90}{NDCG@10}}} & \textbf{\rotatebox{90}{NDCG@100}} & \multicolumn{1}{c|}{\textbf{\rotatebox{90}{P@10} }} & \multicolumn{1}{c|}{\textbf{\rotatebox{90}{NDCG@10}}} & \textbf{\rotatebox{90}{NDCG@100}} & \multicolumn{1}{c|}{\textbf{\rotatebox{90}{P@10} }} & \multicolumn{1}{c|}{\textbf{\rotatebox{90}{NDCG@10}}} & \textbf{\rotatebox{90}{NDCG@100}} & \multicolumn{1}{c|}{\textbf{\rotatebox{90}{P@10} }} & \multicolumn{1}{c|}{\textbf{\rotatebox{90}{NDCG@10}}} & \textbf{\rotatebox{90}{NDCG@100}} \\ \hline
UNSL\#0 & \multicolumn{1}{c|}{0.9} & \multicolumn{1}{c|}{0.81} & 0.63 & \multicolumn{1}{c|}{\textbf{1}} & \multicolumn{1}{c|}{\textbf{1}} & 0.81 & \multicolumn{1}{c|}{\textbf{1}} & \multicolumn{1}{c|}{\textbf{1}} & 0.77 & \multicolumn{1}{c|}{\textbf{1}} & \multicolumn{1}{c|}{\textbf{1}} & 0.76 \\
UNSL\#1 & \multicolumn{1}{c|}{\textbf{1}} & \multicolumn{1}{c|}{\textbf{1}} & \textbf{0.69} & \multicolumn{1}{c|}{\textbf{1}} & \multicolumn{1}{c|}{\textbf{1}} & 0.8 & \multicolumn{1}{c|}{0.9} & \multicolumn{1}{c|}{0.81} & 0.69 & \multicolumn{1}{c|}{0.8} & \multicolumn{1}{c|}{0.88} & 0.72 \\
UNSL\#2 & \multicolumn{1}{c|}{0.4} & \multicolumn{1}{c|}{0.38} & 0.42 & \multicolumn{1}{c|}{0.9} & \multicolumn{1}{c|}{0.92} & 0.71 & \multicolumn{1}{c|}{0.8} & \multicolumn{1}{c|}{0.85} & 0.69 & \multicolumn{1}{c|}{0.8} & \multicolumn{1}{c|}{0.84} & 0.68 \\ \hline
Riewe-Perla\#0 & \multicolumn{1}{c|}{0.5} & \multicolumn{1}{c|}{0.47} & 0.17 & \multicolumn{1}{c|}{0.7} & \multicolumn{1}{c|}{0.62} & 0.74 & \multicolumn{1}{c|}{0.7} & \multicolumn{1}{c|}{0.62} & 0.74 & \multicolumn{1}{c|}{0.7} & \multicolumn{1}{c|}{0.62} & 0.75 \\
NLP-UNED\#1 & \multicolumn{1}{c|}{\textbf{1}} & \multicolumn{1}{c|}{\textbf{1}} & 0.44 & \multicolumn{1}{c|}{\textbf{1}} & \multicolumn{1}{c|}{\textbf{1}} & 0.89 & \multicolumn{1}{c|}{\textbf{1}} & \multicolumn{1}{c|}{\textbf{1}} & \textbf{0.92} & \multicolumn{1}{c|}{\textbf{1}} & \multicolumn{1}{c|}{\textbf{1}} & \textbf{0.92} \\ \hline
\textit{Mean} & \multicolumn{1}{c|}{0.31} & \multicolumn{1}{c|}{0.29} & 0.2 & \multicolumn{1}{c|}{0.33} & \multicolumn{1}{c|}{0.31} & 0.31 & \multicolumn{1}{c|}{0.27} & \multicolumn{1}{c|}{0.26} & 0.26 & \multicolumn{1}{c|}{0.27} & \multicolumn{1}{c|}{0.27} & 0.27 \\
\textit{Median} & \multicolumn{1}{c|}{0.2} & \multicolumn{1}{c|}{0.12} & 0.14 & \multicolumn{1}{c|}{0.2} & \multicolumn{1}{c|}{0.14} & 0.15 & \multicolumn{1}{c|}{0} & \multicolumn{1}{c|}{0} & 0.06 & \multicolumn{1}{c|}{0} & \multicolumn{1}{c|}{0} & 0.7 \\ \hline
\end{tabular}%
}
\end{table}

\section{Conclusion}

In this year's edition, we solved Task 2 related to the early detection of signs of anorexia. The models based on the \emph{CPI+DMC approach} demonstrated robustness when evaluated in a new application domain. 
In turn, the \emph{time-aware approach} allowed including the time during the learning process and optimizing the ERDE$_{50}$ metric, avoiding the necessity of employing more complex decision policies when evaluating the models in an early environment.
This fact encourages us to delve deeper into this approach, as we believe further research is necessary on the combination of precision and speed as a single objective to address ERD problems.


\end{document}